\documentclass[quaderni]{sifrev}
\usepackage{quoting} % L'ho inserito io ma non è coerente con le indicazioni della SIF.
\quotingsetup{font=9pt} %.  Le citazioni sono  in carattere più piccolo.  
 \quotingsetup{font={footnotesize}}

\usepackage{geometry} %per cambiare le dimensioni della pagina

\usepackage{amsmath}           
\usepackage{graphicx}          
\usepackage{rotating} %per ruotare una tabella

\usepackage{hyperref} 

\usepackage[export]{adjustbox}%insieme a \usepackage{array} serve a centrare orizzontalmente nellambiente tabular
\usepackage{array}%insieme a \usepackage{[export]{adjustbox}serve a centrare orizzontalmente nellambiente tabular

\makeatletter
\renewcommand\@makefnmark{%
 \hbox{\@(\textsuperscript{\normalfont\@thefnmark})}}
\makeatother %Nel testo serve ad azzerare lo spazio tra il numero della footnote e la parola alla quale si rifersice 

\usepackage{subcaption}

\usepackage{caption,booktabs}%per centrare o giustificare\caption
\usepackage{xcolor} %per colorare il testo

\usepackage{floatflt}

\usepackage{wrapfig}

\usepackage{mhchem}%per scrivere simbolo elio

\begin{document}

\title{E. Amaldi, C. Dilworth, G.P.S. Occhialini on F.G. Houtermans }

%\shorttitle{Insert a short title here for running head}

\author{Pasquale Tucci} 
% Quick note: Authors must be separated by \ with \inst{n} to mark their referring institute, e.g. \author{A.U. Thor\inst{1} \and W.R. Iter\inst{2}\and S. Einstein\thanks{Thanks here}\inst{3} TeX automagically determines the institute you are referring to.

\institute{retired,  formerly Università degli Studi di Milano}
% Like above you must use \and to separate:
%\institute{Istituto di Fisica Generale Applicata, Universit\`{a} di Milano \and Istituto di Cosmogeofisica del C.N.R - Torino}

\branch{W} %?????
%If gdf: W = one column format L = Letture Scientifiche Q = Questioni Didattiche B = In biblioteca A = Antologia del Giornale di Fisica

\vol{??}
% volume number Aliased in quaderni 

\issue{??}
% issue number

\month{??}
% e.g. \month{Gennaio-Marzo 2011} (gdf) \year{year}  e.g. \month{2011} (quaderni)

%\dedication{Your citation here\\ including paragraphs}   % a complex dedication.

%\riassunto{Nell'Archivio Occhialini-Dilworth dell'Università degli Studi di Milano sono conservati quattro dattiloscritti di Edoardo Amaldi, che trattano della biografia di Houtermans, fisico tedesco, che scelse di trasferirsi in URSS per sfuggire ai nazisti. Durante una purga staliniana fu arrestato e poi consegnato alla Gestapo. I dattiloscritti furono inviati da Amaldi ai due fisici milanesi perché potessero dare un giudizio e suggerire eventuali modifiche. Occhialini e Dilworth, infatti, avevano conosciuto personalmente Houtermans: Occhialini nel 1934 a Cambridge e nel 1950 a Londra; Dilworth a Bruxelles nel 1950 quando scrisse alcune note richieste da Amaldi e inviategli. 

%Lo scenario è quello del ruolo dei fisici nucleari delle due parti opposte nella costruzione della bomba atomica. Lo scenario è stato ampiamente descritto e dibattuto da storici e fisici. In questo articolo verrà analizzato lo scambio di lettere tra Amaldi e i due fisici milanesi al fine di identificare possibili influenze che essi ebbero su Amaldi. In particolare, sottolineerò il contributo di Dilworth, che trova spazio in un capitolo della biografia di Amaldi di Houtermans.} 

\abstract{In the Occhialini-Dilworth Archives of the University of Milan are preserved four typescripts by E. Amaldi, dealing with  F.G. Houtermans, German physicist, who fled to the USSR to escape the Nazis. During a Stalinist purge he was arrested. The typescripts were sent to the two Milanese physicists so that they might give a judgement. G.S. Occhialini and C. Dilworth, in fact, had personally met Houtermans in 1934 and in 1950. The scenario is that of the role of nuclear physicists during the WW2. The exchange of letters between Amaldi and the two Milanese physicists will be analysed in order to identify possible influences on Amaldi. I will highlight Dilworth's contribution, which found room in a chapter of Amaldi's biography of Houtermans.}

\maketitle

\noindent
\hspace{21em}%
\begin{minipage}{0.95\textwidth}
\tiny
%\textit{Usare solo se si intende inserire una citazione}
\end{minipage}

\noindent 
\section{Introducion: Amaldi’s typescripts\protect\footnote{When a person is mentioned for the first time throughout the paper, I give the following information: full name and surname and, in round brackets, year of birth and, if any, year of death. When I'm unable to give the year of death I digit four ``?". When the person is alive I digit four ``.". Thereafter, the person is mentioned only by her/his surname.}}
In the Occhialini-Dilworth Archives\footnote{Occhialini-Dilworth Archives are kept in the BICF, library of Biology, Computer Science, Chemistry, Physics of the University of Milan. For the history of Occhialini-Dilworth Archives see      
\cite{Etra Tucci}} are kept 4 typescripts, as well as several letters, sent to Constance Charlotte Dilworth (1924-2004) (Connie for her friends) and Giuseppe Paolo Stanislao Occhialini (1907-1993) (Beppo for his friends) by Edoardo Amaldi (1908-1989).\footnote{For an historical reconstruction of Occhialini's life see 
\cite{Leonardo Gariboldi}. For Dilworth's life see\cite{Gariboldi Dilworth}.}

The first 155-page typescript, in English, dating back to the early 80s of the 1900,\cite{Amaldi primo dattiloscritto} sent to Dilworth and to Occhialini around the year 1984,\footnote{In the Occhialini-Dilworth Archives I did not find Amaldi's accompanying letter so I’m unable to say when the typescript was sent. I assume that Amaldi sent his typescript around 1984-'85 as the comments of both Occhialini and Dilworth are dated November 1985. } is divided into a Preamble and 22 chapters whose titles are the same as those of the book of E. Amaldi \textit{The adventurous life of Friedrich Georg Houtermans, Physicist (1903-1966)},\footnote{Amaldi could not find a publisher willing to publish his text. Among others, Dilworth also tried to find a publisher after intervening on the text to make it more fluent.\cite{Rufus Neal};\cite{Laurie Mark Brown}} posthumously published for the first time in 1998\cite{Battimelli Paoloni} and in 2012.\cite{Amaldi 1} 
At the end of the typescript are inserted:

\vspace{1.5mm}
\begin{minipage}{0.95\textwidth}
\begin{itemize}
\item[a)]a ``List of publications by F.G. Houtermans (edited by his collaborators of the University of Bern)" featuring 135 items. In the edition published in 2012, 97 items are listed, to which the editors added another 38 items;
\item[b)]
a ``Bibliography" (actually Notes + Bibliography) including 110 items.
\end{itemize}
\end{minipage}
\vspace{2mm}

The second 102-pages typescript, dated ``October 1987", in Einglish, is entitled \textit{The adventurous life of Friedrich Georg Houtermans, physicist (1903-1966)}.
\cite{Amaldi secondo dattiloscritto} The title, the division into chapters and their titles were identical to those of Amaldi's book published in 2012. The List of Houtermans' publications and the Notes + Bibliography are missing. 164 numerical references to footnotes are in the typescript, but their text is missing.\footnote{In the published book in 1998 and 2012 the Notes and Bibliography are inserted at the end of each chapter. In a note Amaldi provided the list of people to whom he had sent the typescript. 
\cite{Amaldi 1}, p. XIII} 

The typescript was annotated by Dilworth who, however, limited herself to the correction of English. In addition she sent by letter to Amaldi some comments.
\cite{Letter of Dilworth to Amaldi 19851107}

The third typescript, identical to the second, is not annotated. It includes the List of publications by Houtermans (135 items) and Notes+Bibliography (89 items). 

The fourth typescript was sent by E. Amaldi on November 23th, 1987.
\cite{Amaldi quarto dattiloscritto} It consisted of 5 sheets ranging from pages 132 to 136 and its title was: ``22 - Why we recall him".  It was the twenty-second chapter with the title ``Why we remember it” of the book posthumously published.  The chapter was sent to two Milanese physicists to check the parts that concerned them, asking specifically Dilworth for integrating it. 

Amaldi, therefore, completed the first draft of Houterman's biography in the early `80s. In a letter sent to Charlotte Riefenstahl (1899-1993) (Schnax for her friends),\footnote{A biography of Riefenstahl, based on many autobiographical material and from other sources, can be found in 
\cite{Shifman Standing}, pp. 30-99.} the first wife of Houtermans,  dated July 29th, 1981, Amaldi attached ``... the draft of the Section 8 to 19 of Fritz Biography".
\cite{Letter of Amaldi to Riefenstahl 19810729}
He continued to work on the Houtermans' biography until at least on October 1987.

The inspiration to write it, as told by the author himself, came from the film ``The Confession", directed by Costa-Gravas, taken from the book by Artur Gerard London ``The confession".
\cite{The confession} Both the film and the book had to deal with the misadventures of London (1915-1986), a member of the Czechoslovak Communist Party, imprisoned in Prague in 1957 on suspicion of deviationism and accused of having contacts with state enemies. Amaldi associated the story of London with that one of Friedrich Georg Houtermans (1903-1966), a German physicist who had experienced similar misadventures in the USSR, where he had voluntarily gone to escape the Nazi dictatorship and to contribute to the construction of a freer and fairer society. Amaldi had known the German physicist personally around 1950.

On November 23th, 1979, Amaldi asked Dilworth for Notes of her conversation with Houtermans in Brussels.
\cite{Letter of Amaldi to Dilworth 19791123} In the letter Amaldi stated that he was writing Houtermans' biography and that he was quite ahead.

Amaldi's book, posthumously published, was based on the draft dated ``October 1987". The editors of the edition of 2012 S. Braccini, A. Ereditato and P. Scampoli added to the text two interviews with Peter Karl Ferdinand Grieder (1928-????) and Johannes Geiss (1926-2020), a  paper by F.G. Houtermans, V. Fomin, L. Shubnikov ``Formation of Heavy Hydrogen Nuclei from Protons and Neutrons" (translated from Ukrainian) and a brief biography of Amaldi. The editors limited themselves to update some biographical data and to insert several images, having been lost the original pictures selected by the author.

Amaldi's book, namely the draft dated “October 1987”, published for the first time by Gianni Battimelli \&  Giovanni Paoloni
\cite{Battimelli Paoloni}.\footnote{In it some notes are incomplete. In the brief introduction to the text, the two editors contextualize the historical as well as the physical interests of Amaldi.
\cite{Battimelli Paoloni}, pp. 475-476 }
  
Amaldi's book, together with that one of Victor Yakovlevich Frenkel (1930-1997), published in Russian in 1997, but largely unnoticed,
\cite{Shifman Preface}, p. X, and in English in 2015\cite{Frenkel english}, are the most complete biographies of the German physicist.\footnote{There are also several other papers that deal in a synthetic way with the misadventures of F.G. Houtermans and with his contributions to the physics. It is worth underlining the paper of Lanfranco Belloni (1944-2017), where several papers and books in Russian on F.G. Houtermans are quoted.\cite{Belloni Houtermans}}
Sources for Houtermans' biography were provided to both Frenkel and Amaldi by Charlotte Riefenstahl. Frenkel, moreover, had the chance to consult archival material, including secret documents, made available by the FSB, the successor of the KGB. The FSB archives were made public only for a short period of time after USSR collapse.\cite{Shifman Introduction}, p. 1 In addition to the KGB archives, consulted thanks to Žores Ivanovič Alfërov (1930-2019), Nobel Prize in Physics in 2000, Frenkel was able to consult material kept in the State Archives of Potsdam, in the former DDR.\cite{Belloni Houtermans}

As Amaldi wrote in the published text ``The adventurous life ...  " Charlotte Riefenstahl, had given him: 

\vspace{1mm}
\begin{minipage}{0.95\textwidth}
\begin{itemize}
\item[a)] 26 pages manuscript she had written for her children about the family of Houtermans; 
\item[b)] ``And he was always right'', typewritten pages dated September 1948; 
\item[c)] 21 typewritten pages written after Houterman's death; 
\item[d)] 7 typewritten pages circulated among a few friends in 1981.\cite{Amaldi 1}, p. 3
\end{itemize}
\end{minipage}
\vspace{1mm}

\section{Who was Houtermans?}
\noindent
Friedrich George Houtermans, Fritz or Fisl for his friends, physicist, lived in a particularly troubled period of world history, experiencing first-hand the Nazi wickedness and the cruelty of Stalin's purges. 

Houtermans was an excellent physicist trained from 1922 to 1927, in the middle of the Weimar republic, in the University of Göttingen, one of the three outstanding centres of German mathematical and physical tradition together with Berlin and Munich.\footnote{A description of Göttingen Scientific environment can be found both in Amaldi's and in Frenkel's biography of Houtermans.\cite{Amaldi 1}, pp. 11-15;\cite{Frenkel english}, pp. 99-106 } At that time the University of Göttingen had in its staff  the mathematicians David Hilbert (1862-1943), Richard Courant (1888-1972), Emmy Noether (1882-1935), Herman Weyl (1885-1955). At the Institute of physics Max Born (1882-1970) had the chair of Theoretical Physics and several younger theoreticians spent months or years working in Göttingen: Werner Heisenberg (1901-1976), Wolfgang Pauli (1900-1958), Enrico Fermi (1901-1954), Giancarlo Wick (1909-1992), Robert Oppenheimer (1904-1967), Paul Adrien Maurice  Dirac (1902-1984), Vladimir Aleksandrovich Fock (1898-1974), Victor Weisskopf (1908-2002), Léon Rosenfeld (1904-1974). 
  
George Gamow (1904-1968) arrived from Berlin in 1928. He and Houtermans wrote a paper inspired by Gamow's paper on the theory of emission of alpha-particles from radioactive nuclei. They collaborated on an extension of Gamow's theory of alpha decay of elements like uranium and thorium, which involved quantum mechanical tunneling or barrier penetration. 

In 1929 Robert d'Escourt Atkinson (1898-1982) (then visiting Göttingen), and Houtermans turned Gamow’s idea around and considered tunneling as a way of assembling complex atoms, thereby arriving at much of the essence of the proton–proton chain put forward in 1939 by Hans Bethe (1908-2005), Nobel Prize 1967 for this work.\cite{Meo} 
At the end of his studies Houtermans got a job at the Technische Hochschule in Berlin-Charlottenburg.  

Also in Berlin Houtermans found a rich scientific environment, ``an impressive collection of scientific talents belonging to four or five different research institutions."\cite{Amaldi 1}, pp. 17-21. In 1929 Atkinson and Houtermans published a more complete and systematic treatment of the theory of the alpha decay of heavy nuclei and used their formula for computing from the experimental values of the energy of the emitted alpha-particle and of the decay constant the radius and height of the potential barrier for each single nucleus of the uranium-radium family.\cite{Amaldi 1}, p. 18

At about the same time they tackled a completely new problem: that of the formation of elements in stars with production of energy. Atkinson and Houtermans computed the probability per unit time of nuclear capture in light nuclei not of alpha particles but of protons under the conditions expected at the centre of a star. For heavy nuclei the probability of
proton capture turned out to be exceedingly small, but for light elements, assuming a collision radius of 4 10\textsuperscript{-13} cm, they obtained half lives ranging from 8 s in  \ce{^{4}_{2}He} to 10\textsuperscript{9} years for \ce{^{10}_{20}Ne}. Then they proceeded to estimate from phenomenological arguments the probability for a proton remaining bound in the nucleus through the emission of its excess energy by radiation.\cite{Amaldi 1}, p. 19

From a detailed analysis of the capture of protons by light nuclei they concluded that processes of this kind can provide the mechanism not only for the formation of light elements, but also for supplying the energy required for maintaining over the ages the emission of radiation from the stars. The authors discussed their paper with Gamow before publication. The results are very close to those given afterwards, as Gamow pointed out some years later.\cite{Amaldi 1}, p. 19

In 1932 Houtermans wrote a paper together with Werner Schulze on the electron microscope. Their paper is considered as the first to have introduced the Louis De Broglie (1892-1987) wave length of the electron as the important physical quantity that should be considered in connection with the resolving power of the electron microscope.\cite{Amaldi 1}, p. 19

A member of the German Communist Party,\cite{Frenkel english}, p. 141 Houtermans fled Germany to escape Nazism and emigrated to England in 1933 with his wife Charlotte and with their daughter Giovanna (Bamsi for their friends). Fisl and Schnax had married few years before, in 1930, with Wolfgang Pauli (1900-1958) and Rudolf Peierls (1907-1995) as witnesses, during a boat trip in the Caucasus organized for the participants to a physics conference in Odessa. 

Charlotte had received her doctorate in physics in 1927 under the physical chemist Gustav Heinrich Tammann (1861-1938) at the University of Göttingen where Houtermans had also received his doctorate in the 1927, under James Frank (1882-1964), Nobel prize in physics for the 1925.

In 1935 Alexandr Leipunskii (1903-1972) invited Houtermans to work at the Ukrainian Physico-Technical Institute in Kharkov, then in the URSS, although some their friends, among them Pauli, had warned not to go. But their hate of the Nazi regime and their hopes on the Soviet regime prevailed. There they hoped to realize their ideals of freedom. 

At the time the Institute was one of the country's leading physics centres and there Houtermans carried out physics researches in Leipunskii's laboratory. Fritz became a good friend of Lev Davidovič Landau (1908-1968) who was used to say: ``If you have a question of nuclear physics, ask Fritz."\cite{Khriplovich Physics Today}, p. 32

But despite his hate for the Nazis, his knowledge of nuclear physics, his ability to collaborate with colleagues, he became a victim of Stalin's purges and, accused of being a spy in the pay of the Nazis, was put in prison in 1937. 
All communications to and from the outside world were interrupted and he was forced to confess to being a Trotskyist plotter and a German spy.\cite{Shifman Preface}, p. XI He hoped that his confession, extracted under torture, would avoid reprisals against his family, who, however, had managed to escape from the USSR.\cite{Shifman Preface}, p. XI

But there is never an end to the worst. As Houtermans himself recounted in \textit{Russian Purge} ... , written by Beck and Godin, \footnote{Beck and Godin are the pesudonims used by Houtermans and Kostantin Feododdovitch Shteppa (1896-1958) to hide themselves. They had been in the USSR prison in the same cell. The real authors were discovered by Frenkel in the 1991.
\cite{Frenkel english}, p. 267. 
Indeed Amaldi had been the first to know the real authors of the \textit{Russian Purge ...}  as he begun to meet Houtermans rather frequently around the 1950.\cite{Amaldi 1}, p. XI. } ``In March [1940] I was ... asked to sign paper ... that I would agree to do secret work for the USSR abroad. This I signed ... otherwise one would be kept indefinitively."\cite{Beck Godin}

Houtermanns was handed over to the Gestapo and put in prison, this time in Germany. Finally on July 16th 1940 he was set free thanks to Max von Laue (1879-1960), Nobel prize winner in physics for the 1914. With the help of von Laue and Carl Friedrich von Weizsäcker (1912-2007), Houtermans obtained a position in January 1941 at a private laboratory directed by Manfred von Ardenne (1907-1997) in Lichterfelde, near Berlin.

Here, in August 1941, Houtermans wrote a report,\footnote{Two pages of the 1941 Report were translated in Italian and you can find them in the Occhialini-Dilworth Archives.\cite{19891115 after} The original paper of Houtermans, in German, was attached by Powers to his letter to Dilworth.\cite{Powers a Dilworth 19891115}} for internal use of the Laboratory, in which he hypothesized the production of energy by means of a chain reaction based on the fission of heavy elements. The paper was published two years later in a magazine of the ``Deutsche Reichpost".\cite{Deutsche Reichpost Magazine} Von Ardenne, in fact, had sent the paper to Wilhelm Ohnesoge, Reich Ministry of Posts. Copies of the report were sent to participants in the Uranverein, launched in 1939: it was attended, at different times, by Walter Bothe (1891-1957), Kurt Diebner (1905-1964), Otto Hahn (1879-1968), Fritz Strassmann (1902-1980), Hans Geiger (1882-1945), C.F. Weizsäcker and W. Heisenberg, along with some others.\cite{Frenkel english}, p. 234

In the report, about 30 typewritten pages long, Houtermans discussed the possibility of producing energy by means of nuclear chain-reactions based on the fission of heavy elements. It was divided in seven sections dealing with:
\vspace{-2mm} 
\begin{quoting}
\item (1) the general point of view; 
\item (2) the processes in competition with fission which can produce an undesired reduction of the neutron density; 
\item (3) the chain reaction based on fission produced by fast neutrons; \item (4) the chain reactions based on the use of thermal neutrons; \item (5) the possibility of realizing a nuclear chain reaction with thermal neutrons; \item (6) the chain reaction in the case of a finite volume of the system;\item (7) the meaning of a chain reaction at low temperature as a source of neutrons and a device for producing nuclear transmutations.
\end{quoting}
The author quoted from the beginning the theory of nuclear fission published in 1939 by Niels Bohr (1885-1962) and John Archibald Wheeler (1911-2008)\cite{Bohr Wheeler} as well as the experimental papers by the Paris and Columbia University groups concerning the emission of secondary neutrons in fission 
\cite{Amaldi 1}, p. 77, footnote 7.

The most interesting points of Houtermans’ paper were, according to Amaldi: 
\vspace{-2mm}
\begin{quoting}
\item a) the conjecture
derived from the theory of nuclear fission that the nuclide of mass number 239
and atomic number Z = 94 (called by its discoverers plutonium: \ce{^{239}_{94}Pu}) should undergo fission even with slow neutrons; 
\item (b) the impossibility of a chain reaction with fast neutrons using ordinary uranium (\ce{^{238}U} 99.3\%;  \ce{^{235}}U 0.7\% );
\item (c) the advantage as a fissionable material of the new nuclide (\ce{^{239}Pu}) with respect to \ce{^{238}U} because the use of the latter requires the separation of this rather rare isotope from the about one hundred times more abundant \ce{^{238}U}; 
\item (d) the possibility of constructing a bomb based on nuclear chain reaction;
\item (e) the possibility of constructing what today is called a fast breeder reactor as the most rational procedure for the exploitation of nuclear energy and the only one which allows the utilization of the whole energy content of natural uranium.
\end{quoting}
Amaldi acknowledged the extreme originality of the paper. If the paper did not produce the effects that could be deduced from it, it is due to the fact that it was not been sent ``to the upper political authorities and the strategic command where it would have triggered a greater interest ... ."\cite{Amaldi 1}, p. 76

The influence that Houtermans' paper on the German atomic bomb program was one of the two points of discussion, after the war, when some physicists posed the problem of Houtermans' rehabilitation from the accusation of having aided the Nazis. 

The other controversial point was Houtermans' return in October 1941 to the Nazi-conquered Kharkov as the representative of the German government in the Luftwaffe uniform-clad.\cite{Frenkel english}, pp. 250-258 As we will see, Houtermans, when in 1950 had a long discussion with Occhialini at Gamow's home\footnote{``My late husband told me that he and Houtermans met for the first time after the war at a conference of the Royal Society in 1950. Then, among others, were invited to the house of a colleague and there they got into a corner and ``had it out" about the Kharkov affair."\cite{Draft of the Letter of Dilworth to Rose after 19950205} } and after with Dilworth in Brussels, denied that he had worn the Nazi uniform. Dilworth quoted Houtermans: ``They will tell you I went to Karkhov in uniform.  It is not true."\cite{Draft Letter Dilworth to Rose 19930722}\footnote{In Frenkel's book many pages are dedicated to Houtermans' return to Kharkov. 
\cite{Frenkel english}, pp. 248-266 The author's goal was to dismiss any suspicion of Houtermans' collaboration with the Nazis. The thesis, based on various documents and testimonies, was that he wanted to save his former colleagues and laboratory equipment. A thesis supported by Occhialini and Dilworth after their meeting with the German physicist in the 1950, as we will see shortly. Amaldi, in his biography of Houtermans, made his own the thesis of those who justified the behaviour of the German physicist.}

\section{Houtermans' meetings with Occhialini}

Occhialini first met Houtermans in 1933 in Cambridge (UK). Houtermans, being a member of the Communist Party, banned after the Reichestag fire of February 27th, 1933 by a fanatic communist, had decided to flee Germany with his family. Hitler had just been appointed chancellor on January 1933. As Occhialini will recall in his Memories written in 1992/'93, he had settled with Houtermans a ``very warm affectionate relationship, based on my unreserved admiration for his brilliance, his political attitude, his knowledge of physics and the originality of his ideas, founded in German culture." 
\cite{Occhialini Memoirs}, p. 66.

In 1938, in one of his periodic returns from Brasil, Occhialini met Charlotte Riefenstahl, the first wife of Houtermans, who told him of the tragedy of the imprisonment of her husband.\cite{Occhialini Memoirs}, pp. 66-67. \footnote{Occhialini's first reaction was to think that it had been his brilliance that had ruined him, that he had talked too much. Houtermans' wife denied all that but Occhialini nevertheless continued to have ``... an uncomfortable sensation, the idea that his wit had brought him to commit indiscretions"}.
 
In that occasion Charlotte asked whether Beppo could meet Frédéric Joliot (1900-1958) in Paris in order to use the influence of the French physicist on Soviet authorities to help her husband. Beppo met Joliot, as the he told in his Memoirs 
\cite{Occhialini Memoirs}, p. 67, and as Charlotte Riefenstahl reported in a letter to Occhialini dated August 5th, 1938: ``Something should come out of this, I have such a feeling as if this project of yours might work. Thanks Peppino!”.\cite{Letter Riefenstahl Occhialini 19380508}\footnote{Riefenstahl's letter, as well as that of July 7th, 1938 and that of December 12th, 1938, are entirely transcribed in the thesis of Mattia Verzeroli, Degree in Physics at the University of Milan, Academic year 2018-2019, Supervisor prof. Leonardo Gariboldi and Co-supervisor prof. Massimo Lazzazoni.}

Indeed Joliot, together with his wife Irène Curie (1897-1956) and Jean Perrin (1870-1942), all three Nobel laureates, sent a telegram to the State Attorney General of USSR and to Stalin. Moreover they sent a long letter to the Attorney General and to Stalin through the Soviet Ambassador in Paris. In the letter they asked the release from prison of Houtermans and Alexander Weissberg (1901-1964)\footnote{Weissberg was the founder of the Soviet Journal of Physics. Like Houtermans he was the victim of a Stalinist purge and suffered a fate very similar to that of Houtermans: imprisoned and handed over to the Gestapo in a prisoner exchange.} on behalf of their “old friendship with the Soviet Union”.\footnote{The telegram and the letter were firstly published in 
\cite{Battimelli Paoloni}, p. 622 and pp. 623-624, and in 
\cite{Amaldi 1}. pp. 46-47. Moreover the text of the telegram and of the letter was sent by Charlotte Riefenstahl to Occhialini in a letter dated July 7th, 1938.}

The letter from Joliot, Curie and Perrin was not followed up favorably, as the relationships between the USSR and the French government had deteriorated. Charlotte Reifensthal wrote to Occhialini in a letter dated December 12th, 1938: ``Are you going to stop at the Collège de France in Paris? Do send me a card in case you are going there. How are you? I have no news, I am very sorry to say."

Only at the end of the war did Occhialini hear of the exchange of Houtermans between the Russians and the Gestapo, and of Houtermans' visit to the laboratory of Kharkov. 

Occhialini met again Houtermans in 1950. Both had been invited to dinner by Gamow. There Occhialini had a long conversation eye to eye with Fisl. At the end of the conversation he was completely convinced. Fisl had changed: in 1933 Beppo had met a teenager but now he had become a man. 
\begin{quoting}
I asked few questions, he explained everything very clearly, his affection for URSS, his indignation for the injustice of the Government who had sent him back to Germany as a Nazi spy, the danger of his situation in Germany and his rescue by von Ardenne and Heisenberg, his brotherly feeling for his Russian colleagues, in particular Leipunskii and his desire to help them which led to his return to Kharkov to help people who might need it. This was the hard point of the discussion. Anti-nazi Germans I had encountered in previous years at Bristol had warned me that Russian colleagues asked about this episode had evaded the question. After this conversation, a dead friend resuscitated had come into my life never more.\cite{Occhialini Memoirs}, p. 67 \end{quoting} 

\noindent Following the clarification Occhialini invited Houtermans to  Brussels to meet Ezra Edgard Picciotto (1921-????)\footnote{Amaldi provided brief biographical informations about Picciotto. Ezra Edgard Picciotto (Italian citizen b. in Istanbul, Turkey, 1921) has studied in Brussels, where he became ``Docteur ès Sciences’’ of the Université Libre in 1952. He made most of his carrier at the same University, except for a number of study periods spent at foreign institutions. His research activity refers to the geochemistry of stable and radioactive isotopes in rocks, ocean and atmosphere. In the period 1957 to 1966, Picciotto concentrated his work on the chemistry of Antarctic ices, taking part in a number of Belgian and American expeditions to the Antartic. He likes to remind that his orientation and formation as a chemist was influenced by Irène Curie and mostly by Giuseppe Occhialini and Friedrich Houtermans.\cite{Amaldi 1} p. 126} and, having to go to  Brazil for a UNESCO mission, he proposed Houtermans to replace him.

At the end of the meeting, however, Occhialini had the impression that the problem of the Kharkov incident remained. ``I feel still the need, not to excuse but to understand him, and also to understand the intelligent people who condemned him.” 
According to Occhialini 
\begin{quoting}
A very witty person is in an unstable equilibrium. He is as it were, a moralist and the intellectual does not forgive a moralist who errs. The wit, who may be called Triboulet\footnote{Nicolas Ferrial (1479-1536) known as Triboulet was  a jester for Louis XII and Francis I. https://en.wikipedia.org/wiki/Triboulet consulted on July 12, 2023} or Oscar Wilde (1854-1900) can be victim of his public when he commits an error. [I already said that the man of 1950 was different from the of 1934.]

\noindent On the other hand although wit has a component of wisdom, when it expressed not in writing but verbally, it can be considered as denoting irresponsibility. One laughs freely at satires on others, not on oneself. I never heard anyone complaining at being hurt by Houthermans wit, but people with Communist sympathies seemed to have found it natural to suppose that a sense of humour tolerated in England should cause him difficulties in USSR. That has been denied by both Fisl and Schnax, who were so alone to the danger of too open speech that they talked of certain things to each other only in bed, under the sheets.\cite{Occhialini Memoirs}, pp. 68-69
\end{quoting}

\section{Houtermans' meeting with Dilworth}

\noindent Occhialini and Dilworth met Houtermans in Brussels, few months after London meeting of Occhialini. This time Dilworth was present. The reasons for the meeting were described by Dilworth in a letter to Amaldi 
\cite{Letter of Dilworth to Amaldi 19800725} who had asked Connie to search through her papers for notes on the conversation she had had with Houtermans in Brussels.\cite{Letter of Amaldi to Dilworth 19791123} Occhialini had informed Belgian colleagues of his willing to invite Houtermans to Brussels, at his research group. But his colleagues were still very sensitive on the subject of war-time collaborations, had heard various rumours about Houtermans: that he had been a Nazi spy in Russia and had collaborated in the German war-effort, and in particular in atomic bomb research. In order to be able to defend him, Beppo started asking questions, lost patience, went off to a cinema and left Connie to carry out a ``third degree".
  In the same letter Dilworth thanked Amaldi for the Touschek article\cite{Letter of Dilworth to Amaldi 19800725}\footnote{Amaldi's Touschek biography is also published in\cite{Battimelli Paoloni}, pp. 505-591.} Moreover she added some verses she had written after Touschek's death.\footnote{
\underline{Bruno Touschek}\\
I met you first in Gottingen, shortly after the war, young, but old in knowing of death and its endless chore. Each for our separate reason, we took the road to Italy to work with welcoming friends, in a spirit that was free.\\
From Rome to Milan without money or map is a long way to go\\ 
In a small car dodging round great lorries caught in the snow.\\
But we met and laughed so often when the wine circled free,\\
We met and mourned to gather friends like Houtermans and Pauli.\\
We grew older, each in his …., and learnt to suffer life,
\\
But you never gave up the struggle, as I did, not bearing strife.
\\
Then I heard you died in Innsbruck, the day that I was there
\\
And I searched your soul in Innsbruck, knowing I must find it there.
\\
You come to die in Innsbruck, home to the land you knew,
\\
(Should I feel death’s breath on my back, I would home to where I grew.)
\\
To a brook that gently falls by sweet flowers, singing
\\
Through green grass waving to the echo of church bells ringing.
\\
Far from the sunburnt hills where the red grape grows,
\\
Far from the bitter joy that the red wine knows.
\\
Softly you laid your head on a pillow of down,
\\
Softly you met your dead, surrendering your crown. 
\\
So to the last you carried a flag with no trace of blood
\\
In a battle without trumpets for a truth not bespattered with mud.
\\
Old Europe died with you, for me and some others,
\\
With the last lingering regrets for a world that called us brothers.}

As Dilworth will recall in a letter to Thomas Powers (1940-....) in 1989, Occhialini had lost patience because in their conversation Houtermans told him that he had deposited, at the post-office of Charlottetown, a claim for a patent for the plutonium idea. ``It was this that made my husband so angry that he went to the cinema and left me to carry on the interrogation. Houtermans assured him that to deposit it at the post-office was equivalent to burying it."\cite{Manuscript of a draft of a Letter of Dilworth to Powers prior to 19891115} Amaldi considered rumors Houternans' deposit, at the Post Office in Berlin-Charlottenburg, of the application for a patent of a fissionable device of military interest.\cite{Amaldi 1}, p. 102 Frenkel didn't mention this rumour in his long discussion of  1941 Houtermans' report.\cite{Frenkel english}, pp. 227-236

Dilworth highlighted, in the letter to Powers, as her report of her ``third degree" with Houtermans was lacking of Houtermans' comments:
 \begin{quoting}This bald summary does not contain the typical Houtermans comment which made his account convincing. For instance, I remember he explained to me that to deposit a report in the Ministry of Posts was perfectly safe, for such was the internal division and petty war between departments in the Nazi bureaucracy, it would never reached the group which was doing real research on the bomb.   
In fact, it appears he was right, but only Fisl could be so sure of his knowledge of the Kafkian complexity of bureaucracy to do such a thing to ensure his scientific priority. This sort of background to the summary is, I am afraid, lost except in my memory. \end{quoting}

The conversation between Houtermans and Dilworth continued, and eventually Dilworth became convinced of Houtermans' good faith. Dilworth had taken notes during the conversation and based on them she wrote 16-point notes. They represent one of the few first and direct witnesses of Houtermans' defense against prosecution about his supposed and controversial collaboration with Nazism.\footnote{Dilworth's Notes
\newline
1. Houtermans was in Gestapo prison in February 1940;\newline
2. He was released the end of July through the intervention of Laue;
\newline
3. Under police supervision as for enemy aliens, not allowed to work in State Institutes or Universities;
\newline 
4. Was found a job by Laue as research physicist in an industrial firm making cyclotrons generators counters etc. under Von Ardenne;
\newline 
5. Police supervision gradually relaxed as for enemy aliens but travelling restricted;
\newline
6. Work at Von Ardenne fully published except one paper;
\newline
7. Spring 1941 Von Ardenne asked Houtermans for a paper on the possibility of a chain reaction; 
\newline 
8. The paper was finished in August 1941, based entirely on published work, mainly on a paper by Zeldovitch and Chariton 1940 in a Russian Journal;
\newline
 9. The content of this paper was:
\newline
a. No chain reaction possible with fats neutrons from ordinary Uranium; 
\newline 
b. Possibility of a thermal pile with carbon beryllium or heavy water;
\newline
c. Influence of Doppler effect on chain reaction;
\newline 
 d. Possibility of a breeding pile producing Plutonium, if the number of neutrons per fission of U is greater than 2, which was not known to Houtermans. This was all highly hypothetical since Houtermans did not know:
\newline 
a. Capture cross section by U; 
\newline
b. Resonance cross section for U, so Houtermans thought that a pile would have to be run at low temperature about —100°C;
\newline
c.  The fact that Plutonium would show thermal fission was highly hypothetical, only derived from Bohr Wheeler considerations;
\newline
10. This paper was given to Von Ardenne who passed it to the Research Department of the Post Ministry, which at that time was not coordinated with the official Uranverein;
\newline
11. Houtermans did not tell his friend Heisenberg about the contents of this paper which remained unread by any competent physicist until after the general coordination of nuclear research in 1944;
\newline
12. Houtermans discussed the paper with two friends Suess and Jensen who were in the Uranium project. He knew them to be anti-Nazi, having been introduced to Jensen by Rompe his friend since 1932;
\newline
13. Suess is mentioned by initial as having assisted the Allies by giving information to the Norwegian heavy water engineers in a book by one of them about the role of the heavy water plant in the war. This can be checked;
\newline
14. In automn 1941 Houtermans asked Weizsäcker whom he took at that time to be a anti-nazi whom he knew to be going to visit Bohr to tell Bohr the whole position of the Uranium project in Germany and to get absolution for the part played by German physicists in nuclear research since there was no danger of it being used in the war;
\newline
15. Neither Weizsäcker nor Heisenberg did so;
\newline
16. Houtermans hearing of this through Suess who had met Moller, commissioned Jensen to do the same on behalf of himself Suess and Houtermans on his visit to Bohr in 1942.\cite{Letter of Dilworth to Amaldi 19800725}}

The points  8., 9., 12., 14., deserve attention.\\
In the point 8. Houtermans, as Dilworth reported, based his paper on that one of two Russian scientists: a) Yakov Borisovich Zeldovich (1914-1987) played an important role in the development of Soviet nuclear and thermonuclear weapons. b) Yulii Borisovich Khariton (1904-1996) directed the Soviet nuclear program for many years.\cite{Frenkel english}, pp. 222-223
On the contribution of Zeldovich and Khariton to nuclear physics see\cite{Ostriker Zeldovich}.

In the point 9. Dilworth described the content of Houtermans' article ``Zur Frage ... "\cite{Deutsche Reichpost Magazine}. In it Dilworth pointed out that the paper was based on published work and was highly hypothetical because Houtermans didn't know 

a. the capture cross section by U;

b. the resonanace cross section by U;

c. the fact that Plutonium would show thermal fission was highly hypothetical, only derived from Bohr Wheeler considerations.  

Dilworth argued that the paper would have been of little practical use, giving substance to the thesis that Houtermans did not give any aid to the German war effort.

In the point 12. Dilworth wrote that Houtermans had communicated the contents of his paper to Hans E. Suess (1909-1993)\footnote{Suess was an expert on heavy water and a scientific advisor to Norsk Hydro, the Norwegian plant in Vemork, producing hydrogen by electrolyzing water  and as a by-product of heavy water.\cite{Waenke Suess}} and Hans Jensen (1907-1973)\footnote{He worked with Paul Harteck (1902-1985) on heavy water. In the 1942 he talked to Bohr about his researches in Germany.} who were in the Uranium project and that he knew them as anti-Nazi, having been introduced to Jensen by Robert Rompe (1905-1993), his friend since 1932.

In the point 14. Dilworth reported that in autumn 1941 Houtermans had asked Weizsäcker, considered at that time to be a anti-Nazi, whom he knew to be going to visit Bohr, to tell to Danish physicist the whole position of the Uranium project in Germany and to get absolution for the role played by German physicists in nuclear research since there was no danger of it being used in the war.

Ultimately, Dilworth's impression from the conversation with Houtermans about his 1941 paper was one of complete innocence and that his anti-Nazism was fully proven.

\begin{quoting}
\noindent I do not know what transpired these discussions, but Fisl remained in Brussels and started a very positive collaboration with Picciotto on “plombiologia”, i.e. geophysical dating by means of lead isotopes. Later, he, Fisl, was offered a chair in the University of Berne, where he created an extremely vital centre on geophysical data, carried on by his pupil Geiss.\cite{Draft of a Letter of Dilworth to Powers prior to 19890418}
\end{quoting}

While the points taken during the conversation were lost, the Notes were found by Dilworth and were sent to Amaldi in 1980.\cite{Letter of Dilworth to Amaldi 19800725}

She sent her Notes also to Thomas Powers, journalist, who was writing a book about Heisenberg.\cite{Powers Heisenberg} with the premise of taking ``into account my age - at the time $\sim$ 26 years - and my consequent ingenuity.”\cite{Draft of a Letter of Dilworth to Powers prior to 19890418}

\noindent
\section{Occhialini's comments to Amaldi's Biography of Houtermans}

Occhialini annotated the first Amaldi's typescript in several parts. In general, Beppo's aim was to highlight or confirm Amaldi's statements. Few of them, however, - they can be counted on the fingers of one hand - showed different points of view. 

For example, in chapter 13 entitled ``A few other physicist’s political troubles" Amaldi recalled the story of Pjotr Leonidovich Kapitza (1894-1984) whom he and Emilio Segré (1905-1989) had met in 1934 at the Cavendish and Mond Laboratories in Cambridge (UK). As it is known, when in 1934 Kapitza went to the USSR to participate in the Mendelee'ff Congress, organized on the occasion of the centenary of the Russian chemist, he did not receive the passport for the return and the direction of a new laboratory was offered to him.
Amaldi wrote:
\begin{quoting}
When, in 1978, Leonidovich Kapitza was awarded the Nobel Prize for his discovery of superfluidity, all physicists throughout the world were very happy to see the recognition of his outstanding piece of work which emerges from a full life devoted, with great success, to the progress of science. 
But everybody that had the great luck of meeting him or at least of knowing some detail about his life, was also satisfied or even moved, that such a recognition was given to a person endowed of such extraordinary values.\cite{Amaldi primo dattiloscritto}
\end{quoting}

\noindent Occhialini underlined ``extraordinary human values" and added:
\begin{quoting}
Era difficile ma la trasposizione del periodo 1934 dopo il 1937 è faticoso.\footnote{Translation: It was difficult but the transposition of the period 1934 after 1937 is tiring.} See also his behavior at the Sakharov trial at the USSR Academy.
\end{quoting}
Occhialini was referring to the following episode: when in 1980\footnote{The Decree of the Presidium of the USSR Supreme Soviet is dated January 8th, 1980.} Andrej Sakharov (1921-1989) was stripped of the title of Hero of Socialist Labor and of all his state awards and exiled, from 1980 to 1986, to Gorki (now Nizhny Novgorod), a city then off limits to foreigners by decision of the authorities, Kapitsa was asked to intervene but, as Sakharov wrote in his Memoirs, `` ... like our other ``celebrities", he did nothing."\cite{Sakharov Memoirs}, p. 517,\footnote{A. Marchenko wrote ``An open letter to Acedemician P.L. Kapitsa" dated March 1, 1980 asking him an active intervention in defense of Sakharov, but even this appeal was not followed up.\cite{open letter}, pp. 31-37.}

On the other hand, Sakharov credited Kapitsa with not having signed, in 1973, a letter in which about fifty academics accused him of having carried out actions alien to Soviet scientists.\cite{Sakharov Memoirs}, p. 303, p. 517. 

There is again a long comment by both Occhialini and Dilworth on Houtermans' moving letter to his mother published by Amaldi in chapter 14 entitled ``Finally out of prison". 
Dilworth commented: “Sarebbe certo interessante di conoscere più lettere di Fisl che certamente esistono e che potrebbero trovare posto in una biografia. C. D.”\footnote{Translation: It would certainly be interesting to know more letters of Fisl that certainly exist and that could find a place in a biography. C. D. (My transaltion P.T.)}

Occhialini wrote: 
\begin{quoting}
La lettera prova qualche cosa di più che l’attaccamento alla famiglia. È un documento culturale impressionante e vorrei vedere le altre perché costituiscono la definizione del secondo Houtermans che mi ha colpito dopo la guerra per la maturazione prodotta dalla prigionia (Heine again (protestante + ebreo)). Mostra come il ``wit e il senso of humour” vengono dal profondo e fa desiderare di leggere ancora di queste lettere con una traduzione inglese perfetta. Da notare l’assenza di self pity.\footnote{Translation: The letter shows something more than attachment to family. It is an impressive cultural document and I would like to see the others because they constitute the definition of the second Houtermans that struck me after the war for the maturation produced by imprisonment (Heine again (Protestant + Jew)). It shows how the ``wit and sense of humour" come from deep inside and makes you want to read more of these letters with a perfect English translation. Note the absence of self pity.(My transaltion P.T.)}\end{quoting}

Amaldi left the chapter unchanged in the final published version of Houtermans' biography. It is very likely that the copy commented by Occhialini was not sent to Amaldi. 
Instead, Amaldi received from Occhialini two handwritten sheets with various considerations.\cite{Letter of Occhialini to Amaldi 19851108}
\begin{quoting}
For he who lives more than one life more than one death must die" (O.W., citato a Memoria, prob[abilmente] sbagliato.\footnote{Wilde: ``For he who lives than one / More deaths than one must die."\cite{Wilde}}
\end{quoting}
After listing the small deaths of Houtermans Occhialini continued:
\begin{quoting}
La crisi più grave è la $3^a$ come ho capito una sera a Londra ($\sim$ 1950) che ha marcato la nostra riconciliazione (non c'eravamo incontrati più dal 1934). Non è stata da parte mia un adattamento, sono rimasto convinto, anche perchè ricordavo (Schnax 31 Dic 1938 - 1 Gennaio in Londra) che il solo fatto di essere stato arrestato era un peccato per la sinistra europea.\footnote{The most serious crisis is the $3^a$ as I understood one evening in London ($\sim$ 1950) that marked our reconciliation (we had not met since 1934). It was not an adaptation on my part, I remained convinced, also because I remembered (Schnax 31 Dec 1938 - 1 January in London) that the mere fact of being arrested was a shame for the European left. (My transaltion P.T.)}
\end{quoting}
Two notes concerned Houtermans' return to Kharkov. Fisl denied that he went into Nazi uniforms and that he had a military cap.

\section{Dilworth's comments to Amaldi's biography of Houtermans}

Dilworth listed 6 points where she disagreed with Amaldi.
\cite{Letter of Dilworth to Amaldi 19851107} The fifth point was interesting. Dilworth argued that Fisl's security that his paper of 1941 would remain buried stemmed from his theory of  ``Baumaschology", according to which an official document can be used for any end except the one for which it was issued. And he gave the example of a free pass for a train journey which will not accepted by the station ticket office but may be used to get petrol off the ration. 

The most important part of Dilworth's considerations, however, dealt with the question why we remember Houtermans. It is worth quoting this part in full because it was almost entirely inserted by Amaldi in the chapter ``Why we recall him''.
\begin{quoting}
As an individual but also as a representative of a culture that is lost. The latter is what he had in common with Bruno Tousheck (1921-1978).\footnote{Amaldi published a Touschek's biography two years after the death of the Austrian scientist. 
\cite{Amaldi Touschek 1981}. Amaldi's Touschek biography is also published in\cite{Battimelli Paoloni}, pp. 505-591.} Both were essentially 'Mittel Europa', Germanic in the best sense, of the anarchist breed; in revolt against the Prussian element, and against bureaucracy. Individualist, romantic, but with a crystalline cynicism with regard to any form of fanaticism: enthusiastic and in love with life, but life of the city. You cannot imagine either of them planting cabbages. That both were hard drinkers is indicative of the melancholy that underlies ebullience.

Professionally, both were much better than their work, their intelligence was at least equal to the most successful of the contemporaries. The lack of success was in part due to being in the wrong place at the wrong time.

Houtermans was an innocent, of the tribe of Peter Pan. His refusal, to grow up, to become serious, was the basis of his charm. On the other hand, as all Peter pans, he never understood women. His four marriages, three wives and many children do not denote the profligate but rather a child seeking the road to the Never Never Land.

Part of this was his warmth and generosity. He may have lied, betrayed and twisted truth to feed his vanity, but he was not mean. He had no ideal, but was idealistic in the sense that he did not live for today or tomorrow. He lived in the stream of history, egoistic but not egocentric, that his culture forbade. Above all, life was a great joke, a continuous laugh against the lumpen-bourgeoisie. 

They don't make them like that anymore. That's why we remember him.\cite{Letter of Dilworth to Amaldi 19851107}
\end{quoting}

Amaldi pointed out that the letter that had been sent  by Connie was very interesting and he wanted to quote in full a long passage contained in it.\cite{Letter of Amaldi to Dilworth 19871123}

Both the content and the date of Connie's letter were confirmed by a Beppo's letter to Amaldi of November 8, 1985, in Italian.\cite{Letter of Occhialini to Amaldi 19851108} Beppo wrote that he went to Rome for a meeting of GIFCO (Gruppo Italiano di Fisica Cosmica) and wanted to personally give Connie's synthetic notes on Houtermans, but Amaldi was not in the Institute.
In addition to being interesting for the clarification, the letter is moving because in it Beppo confessed his melancholy: 
\begin{quoting}
Ho vissuto uno di questi periodi crepuscolari. Gli ultimi mesi sono saturi di scomparse di amici, viaggi per tentare di arrivare in tempo a salutarli per l’ultima volta, visita e lettere per tentare di consolare famigliari, notti di incubi e rimorsi.

\noindent Sto emergendo dalla crisi e nelle prossime settimane ti manderò l’estratto di varie dozzine di pagine di divagazioni sopra Fisl. 

\noindent Se, a causa del ritardo non c’è + posto fammelo sapere con dolcezza.
\footnote{``I lived one of these crepuscolar periods. The last few months have been saturated with disappearances of friends, trips to try to arrive in time to say goodbye for the last time, visits and letters to try to comfort family, nights of nightmares and remorse.\\
\noindent I am emerging from the crisis and in the coming weeks I will send you the extract of several dozen pages of digressions above Fisl. \\
\noindent If, due to the delay there is no + place let me know sweetly." (My translation. P.T.)}
\end{quoting}
 
Occhialini expressed the same mood in a letter to his brother in 1981: ``Milano mi fa paura, e Genova è per me troppo triste."\footnote{``Milan scares me, and Genoa is too sad for me." (My translation. P.T.)} Remembering the death of his friend Touschek he wrote to Amaldi: 

\begin{quoting}
... questa morte risuscitando le memorie di Pauli e di Houtermans ha chiuso tanto del passato della fisica che sono rimasto con gli occhi che volevano guardare altrove. Mi sono concentrato nel futuro di questo pezzo di terra [la sua casa in Toscana] cercando di scacciare i fantasmi.
\footnote{``... this death resurrecting the memories of  Pauli and  Houtermans closed  so much of  the  past  of physics that  I remained with eyes  that wanted to look elsewhere." (My translation. P.T.)}\end{quoting}
  
Quotations were reported by Agnese Mandrino in the Foreword to the Inventory of the Occhialini-Dilworth Archives compiled in the years 1998-2000.\cite{Mandrino inventario}

In 1992 Dilworth, probably under the pressure of writing the Memoirs,\cite{Occhialini Memoirs} came back to Houtermans' biography of Amaldi. In the Archives Occhialini-Dilworth seven unpublished sheets of Dilworth's manuscript are kept.\cite{Dilworth seven sheets undated after 1992} The sheets deal with different themes and some sheets have been lost. So some Dilworth's considerations are interrupted. Nevertheless, they are very interesting because they offer us Dilworth's thought in all its complexity. They are undated, but a reference to Giorgio Salvini (1920-2015)\cite{Salvini Amaldi} places the manuscripts after the publication date of Salvini's paper. In them Dilworth aimed to analyze what Amaldi had written on Touschek and Houtermans splitting her reflections in three points:\\
1. Critique of Amaldi Method;\\
2. Explanation of re-presentation;\\
3. Historical background.

As regards to the point 1. Dilworth wrote that Amaldi was not an historian and presented mostly personal memoirs with little verification from documents and rarely confronted with other testimonials. But this could not be considered a defect because it had the advantage of vivacity, ``... lightening Amaldi's pedantic style. As a physicist he couples to the personal drama, long explanations of the personal activity. This is valid approach in that necessary but indigestible to the common reader.\cite{Dilworth seven sheets undated after 1992}, sheet 1
 
As regards to the point 2. Dilworth highlighted how in order to overcome the difficulties of the original text the editor had taken some liberties with it.\\ 
\indent As regards to the point 3. Dilworth remembered that once when they were talking about his essay on Houtermans she asked Amaldi why he choose two such subjects, Touschek and Houtermans (at that time Dilworth was not aware of Amaldi's essay on Majorana). \\
\indent Amaldi replied that it was because he did not understand them. According to Dilworth, that he did not, is part of the fascination of these attempts. Here she quoted Salvini, who remembered Amaldi's ``exemplary cold rationality" and Dilworth commented:
\begin{quoting}
This puzzled rationality, seeking to understand intuitive imaginative characters is the essential theme of this book. It underscores the complexity of characters attracted to what the general public is led to believe as the cold objective attitude to science. In truth, nothing is rarer in the profession. The personality of Amaldi which comes through his writings is far from typical.\cite{Dilworth seven sheets undated after 1992}, sheet 2
\end{quoting}
What did Majorana, Houtermans and Touscek have in common? Why Amaldi wrote their biography? According to Dilworth the three physicist had in common the passion for their work.  And it was the very passion which destroyed them, the same passion which helped Houtermans to survive four years of prison of the KGB.\footnote{Indeed Dilworth claimed that Houtermans was in Soviet prison for seven years. But this is an imprecise recollection.}
\begin{quoting}In Amaldi’s careful, objective, (though uncritical) account of the lives of his personages there are tantalising glimpses of a Europe which is being fast forgotten (Houtermans, in post-world-war I was part of the Adler group (so also was the wife of Pauli, Franca). It was the time of the Weimar Republic, of the``Berlin Diaries of Christopher Isherwood (1904-1986)\footnote{Berlin is portrayed by Isherwood during this chaotic interwar period as a carnival of debauchery and despair inhabited by desperate people who are unaware of the national catastrophe that awaits them. 
$https://en.wikipedia.org/wiki/The_Berlin_Stories
 consulted in July 11th, 2023$} and in USA of the tragedy of Sacco e Vanzetti.\cite{Dilworth seven sheets undated after 1992}, sheet 6 \end{quoting}

Finally, Dilworth emphasized the ``esprit de corps" of the scientific community: Von Laue helped Houermans to find a job after his imprisonment in Russia and Germany. 

Arnold Sommerfeld (1868-1951) helped Touschek to move from Vienna to Hamburg where Jensen and Wilhelm Lenz (1888-1957), strictly related to Sommerfeld, could be trusted in supporting Touschek in his semi-clandestine life in Germany, where his Jewish origin was not as easily detectable as it had been in Vienna.\cite{Bonolis Pancheri Touschek} 

Perrin and Joliot, at the time of their action in support of Houtermans were not in danger, but still their letters showed that the community took its responsibilities very seriously.

\noindent

\section{Conclusions}
Unlike Amaldi who wrote a book with the goal of reconstructing historically the adventurous life of Houtermans, Dilworth and Occhialini didn't have such a goal, although Dilworth claimed to have ``dabbled in History of Science" after her retirement.\cite{Draft of a letter of Dilworth to Powers undated after 19890418} Considerations of the two Milanese physicists on Houtermans, in fact, are scattered through a wide range of notes, not intended for printing, and in private letters.

Nevertheless it is possible to identify quite clearly three different ways of analysing the figure of Houtermans, although they had in common the intent to free Houtermans from the shameful accusation of having helped the German war effort during World War II.

In a manuscript Dilworth reported that she had asked Amaldi why he had taken an interest in Houtermans and Touschek, the Roman physicist replied ``because I did not understand them''.  And immediately afterwards Dilworth reported Salvini's judgment on  rational behavior of Amaldi who, evidently, judged Houtermans' behavior irrational. Amaldi observed in Houtermns two completely contradictory personalities: from on the one hand the physical and objective side  of which Amaldi appreciated his numerous and profound papers; on the other hand, the subjective side with behaviours incomprehensible to the Roman physicist: Houtermans had married four times and twice with the same person; the first time he had suddenly married Charlotte during a boat trip organized during a conference in Odessa; he had gone to Russia despite the misgivings of many of his friends including Pauli; he had returned to Kharkov in Nazi uniform ans so on.
In general, Amaldi also found incomprehensible that a good physicist, in a valid institution as the Ukrainian Physico-Technical Instritute in Kharkov, in a nation chosen because Houtermans hated Nazism, could be persecuted.

Dilworth and Occhialini, on the other hand, appreciated Houtermans as physicist but they tried to understand Houtermans' personality. 

Occhialini, as well as Dilworth, didn't want to reconstruct historically the events of Houtermans' life. Beppo was convinced of Houtermns' innocence not on the basis of documentation but on the basis of what Houtermans told him as justification in the long meeting at Gamow's house. Occhialini felt that Houtermans had been too wit and that if this characteristic was appreciated in Germany or England it could create problems for him in Russia, although in conversation with Charlotte Riefenstahl, Houterman's first and third wife, she had excluded this possibility.

According to Occhialini wit has a component of wisdom, when it expressed not in writing but verbally, it can be considered as denoting irresponsibility. People with Communist sympathies found it natural to suppose that a sense of humour tolerated in England should cause him difficulties in USSR. 

However, Occhialini's proposal to offer Houtermans the possibility of replacing him on the occasion of his trip to Brazil on behalf of UNESCO was decisive for Houtermans' complete rehabilitation. And no less decisive for Houtermans was Occhialini's ability to convince his fellow physicists of Houtermans' innocense.

Dilworth claimed that Houtermans was an eternal Peter Pan. His refusal, to grow up, to become serious, was the basis of his charm. And while Occhialini claimed that between 1934, the date of their first meeting, and 1950 Houtermans had become an adult, for Dilworth, instead, Houtermans had remained an eternal child. 

According to Dilworth, as all Peter pans, he never understood women. His four marriages, three wives and many children do not denote the profligate but rather a child seeking the road to the Never Never Land.

However according to Dilworth he may have lied, betrayed and twisted truth to feed his vanity, but he was not mean. Life was a great joke, a continuous laugh against the lumpen-bourgeoisie. 

Dilworth pointed out that individuals such as Houtermans or Touscheck were the last specimens of a species that died together with the Weimar Republic. 

Houtermans' strong passion for physics, and science in general, allowed Houtermans to withstand four years of imprisonment. But Houterman as all brilliant physicists, who strongly claim the priority of an important discovery, in 1941 handed von Ardenne a document that could have terrible consequences. 

But Dilworth, not forgetting that she was a physicist and, analysing the physical content of the 1941 report, drew the conclusion that it could not have practical consequences, since too much relevant data was missing. Although she admitted that in the 1941 report Houtermans spoke explicitly of a war use of a possible product of his research.

Dilworth and Occhialini were prompted by Amaldi to rethink the Houtermans affair as a couple decades after they had played an important role in dispelling any doubts about Houtermans' innocence.

In conclusion, a conviction is common to the three interpretations of Houtermans: his innocence.

\section*{Acknowledgment}
I warmly thank Doct. Daniela Martella, Librarian of the BICF, Università degli Studi di Milano, and Doct. Flavia Safina, BICF, for their assistance in consulting the Occhialini-Dilworth Archives as well as Doct. Antonella Cotugno, Archivist and Librarian of the Library and Archive of the Physics department of the University of Rome, for her assistance in consulting Amaldi Archives.\\
A very special thanks goes to Prof. Gianni Battimelli who was of fundamental help in guiding me among the sheets and the papers of Amaldi Archives.

 \renewcommand{\bibname}{BIBLIOGRAPHY}

\end{document}